\documentclass[11pt]{article}
\usepackage{authblk}
\usepackage{graphicx}
\usepackage{xcolor}
\usepackage{natbib}
\usepackage[margin=2cm]{geometry}

\title{\bf ESO Expanding Horizons White Paper:\\
Habitability of exoplanets orbiting flaring stars\footnote{This is 
a white paper submitted in response to the ESO Expanding Horizons 
initiative.}}

\author[1]{Rebecca Szab\'o}
\author[2]{Valentin D. Ivanov}
\author[1]{Michal \v{S}vanda}

\affil[1]{Astronomick\'y \'ustav Univerzity Karlovy, Czech Republic}
\affil[2]{European Southern Observatory (ESO), Germany}

\date{December 2025}

\begin{document}
\maketitle

\begin{abstract}
As of late 2025 there are about 70 exoplanets that meet the formal 
criterion of having equilibrium temperatures allowing the presence of 
liquid water and about 50 of them orbit M-stars, known for their strong 
chromospheric activity. Most of these stars are close to the Sun and 
the planet-to-star mass and luminosity ratios are advantageous, allowing 
for a more detailed follow-up than of planets orbiting hotter and more 
massive stars. Many more planets orbiting late-type stars are expected 
to be discovered by {\it Gaia} and {\it Plato} in the following years.
However, the lingering question remains whether the UV and X-ray emission,
associated with the stellar activity, allows for complex life. A 
comprehensive study focused on properties of flaring exoplanet hosts and 
their activity, on a much larger scale than these few tens (soon to 
become hundreds) of stars with habitable planets is called for, to answer 
the question if such stars can harbor habitable planets. The proposed 
Wide Field Survey telescope is well suited for this study.
\end{abstract}

\section{Exoplanet Habitability}

The issue of exoplanet habitability
became much more tangible after the discovery of the first exoplanet 
orbiting a solar-type star \citep{1995Natur.378..355M}. The habitable 
zone (HZ) is defined as the range of orbital radii where the planet's 
equilibrium temperature allows liquid water to exist on the surface. 
This is a loose definition: planet masses, presence of the greenhouse 
effect, geology, orbital eccentricities, and possibly other factors 
are expected to play a role in determining the HZ borders. The most 
reliable catalog of HZ rocky exoplanets today lists 67 objects 
\citep[typical radii $\sim$1.5\,R$_{\mathrm{Earth}}$, 
masses 1--3\,M$_{\mathrm{Earth}}$ -- ][]{2025arXiv250114054B}. 
This list is expected to grow with the next {\it Gaia} data release 
and future transit missions like {\it Plato}.

Finding true Earth-Sun analogs requires multi-year monitoring. The 
HZ is closer in and orbital periods are shorter ($<$100\,days) for 
lower mass stars, making the discovery of such planets easier. M-dwarfs 
make up over 60\% of stars within 10\,pc \citep{2021A&A...650A.201R}.
Their evolution is slower than for earlier-type stars, allowing a more 
stable HZ \citep{2025ApJS..281...13L}. These considerations 
attracted attention to stars cooler than the Sun, but it was 
quickly realized that their activity is a habitability risk -- 
close-in exoplanets are more exposed to intense flares and coronal 
mass ejections
\citep{2024ApJ...960...62E,2024ApJ...966...69L} that can photoevaporate
their atmospheres \citep{2023MNRAS.525.5168M}.
\citet{2010AsBio..10..751S} argued that the flares can affect the 
exo-atmosphere chemistry, e.g., causing years-long ozone depletion. 
Fortuitously, the UV environment of quiet M-stars may be even lower 
than the Sun's \citep[see for a review][]{2013Sci...340..577S}, but 
not all of them are quiet; furthermore, stellar activity declines 
with age \citep{2024ApJ...960...62E}.

\section{Stellar Flares}

Stellar flares
represent a complex of phenomena during which a rapid release of 
energy occurs due to the reconnection of the magnetic field. The 
first flare was observed visually on the Sun \citep[][the mightiest 
in modern history]{1859MNRAS..20...13C}. The advance of 
observational techniques in the 20$^{\mathrm{th}}$ century showed 
that flares can be captured quite often. The energy release during 
a typical flare is always accompanied with an emission of EUV and X-ray 
electromagnetic radiation, typically also accompanied by the 
eruption of hot solar plasma confined in a magnetic field -- the 
coronal mass ejection. Soon it was realized that the Sun is not 
unique in this regard, and superflares -- much more powerful 
equivalents of what has ever been seen on the Sun -- may be 
observed on other stars. 

Until the seminal paper by \citet{2012Natur.485..478M} it was 
believed that superflares occur either on stars of different spectral 
type or evolutionary stage, or on stars in interacting systems. 
This work showed that even single solar-like stars exhibits superflares, 
prompting questions like how these solar-like stars differ from our Sun 
so that we do not observe superflares on our Sun, if it is possible 
that the Sun exhibits a superflare occasionally and what the effects 
of such events on life on Earth would be. 
Almost in parallel, \citet{2012Natur.486..240M} reported on finding 
indications of much stronger than Carrington-event occurring in the 
historical era. Until today, several such Miyake-events were reported 
in the literature \citep[see for a review][]{2023SSRv..219...73U},
raising the question about the interaction of flares and planetary 
biospheres once again. 

\section{Planetary Environments and the Potential for Life}

Impact of stellar activity 
on planetary environments and the potential for life require 
accurate estimates of flare energies. Radiation and particle outputs 
profoundly influence planetary atmospheres. \citet{2019AsBio..19...64T} 
showed that high-energy flares at $\sim$$10^{34}$\,erg occurring at a 
frequency of at least once per month can deplete over 99.99\% of a 
planet’s ozone layer, exposing the surface to potentially sterilizing 
levels of UV. They estimated that about 8.3\% of such flares from 
M-dwarfs would directly impact a planet located in the HZ, posing 
serious risks to atmospheric retention and surface habitability. 
\citet{2019ApJ...873...97L} estimated that flares with energies 
$>$10$^{34}$~\,erg occur about once every 6.5\,yr. Even under more 
conservative assumptions, flares could prevent the long-term survival 
of Earth-like ozone layers. Beyond atmospheric erosion, flare energy 
may play an important role in prebiotic chemistry. 

\citet{2018SciA....4.3302R} defined {\bf abiogenesis zones} as regions 
around stars where UV-driven chemical pathways crucial for life’s 
origin can occur. These zones are constrained by the competition 
between productive photochemical reactions (e.g., those forming 
ribonucleotides essential for RNA synthesis) and side reactions that 
lead to inert by-products. The key factor in sustaining these
productive reactions is the cumulative UV energy delivered e.g., by 
stellar flares, particularly in the 200--280\,nm band. The authors 
calculated the required UV flux accumulated over the lifetime of 
intermediate compounds, demonstrating that only planets receiving 
sufficient UV flare output can support this form of prebiotic 
chemistry. Flare energies and flare occurrence rates together with 
their spectral energy distribution are the key to estimating the 
possibility of the abiogenesis zone to be present around a particular 
star. This idea was further processed by \citet{2020AJ....159...60G}. 
On the other hand, proton events often associated with flares may 
have adverse effects on life on the surface of the nearby planet. 
Even though \citet{2017MNRAS.465L..34A} estimated that a complete planetary
surface sterilization is unlikely by stellar proton events, 
severe effects on possible planetary biospheres can still be 
expected, possibly even in the form of large-scale extinction events. 

\section{Solar Flares}

The study of Solar 
flares has advanced significantly over the recent years. Their 
near-continuous monitoring is traditionally performed in 
chromospheric lines (e.g., H$\alpha$, but not exclusively in 
those), with simultaneous soft X-ray space-based monitoring from 
the GOES satellite, and in the last decades also in UV, with e.g., 
EIT/SOHO or AIA/SDO. Radio observations from Earth have also 
been very useful in continuous monitoring. The X-ray and radio 
observations are mostly integrated over the entire solar disk. 

Spectroscopic observations of solar flares have been obtained 
primarily in the course of spectroscopic monitoring campaigns.
The available material is surprisingly abundant \citep{2024ARA&A..62..437F}. 
Therefore, the spectroscopic properties of solar flares seem to be
reasonably well understood. 

\begin{figure}
\centering
\includegraphics[width=1.0\textwidth]{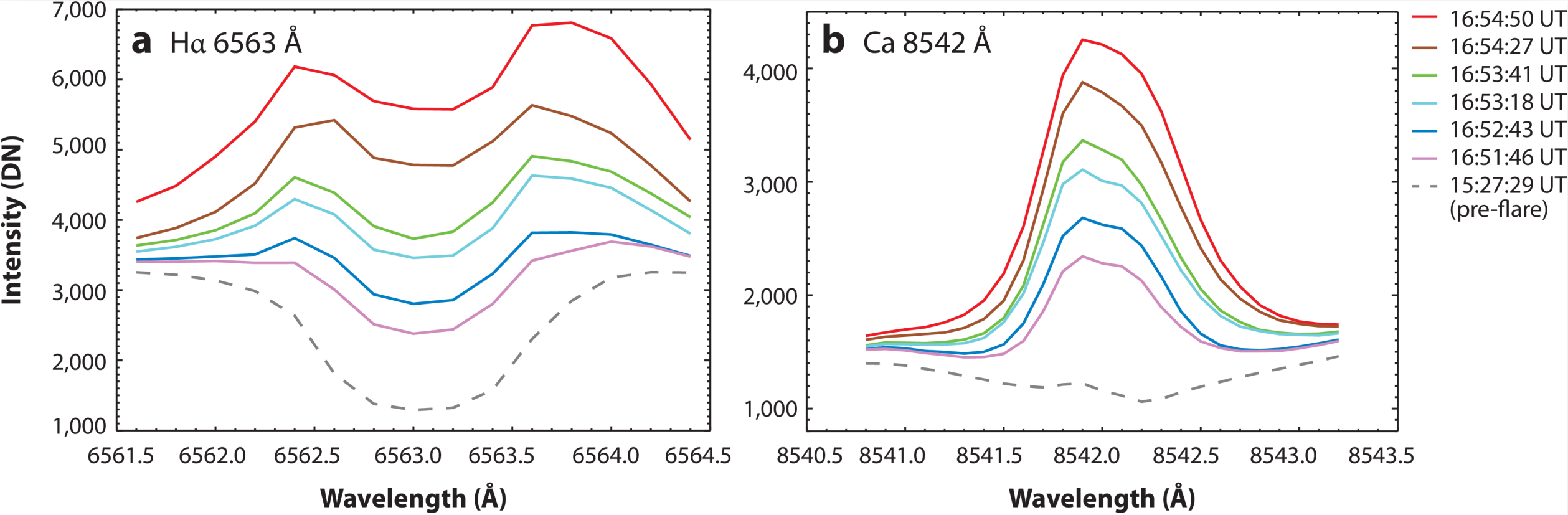}
\caption{Temporal evolution of H$\alpha$ and Ca\,8542 in a solar flare 
spectrum \protect{\citep{2015ApJ...813..125K}}. The duration of this 
flare is of the order of minutes. Some flares may continue for tens of 
minutes to hours.}\label{fig1}
\end{figure}

On the other hand, spectroscopic information about flares on stars 
other than the Sun (see Fig.\,\ref{fig1}) is sparse. Several 
flare stars 
are continuously being monitored within amateur networks, where the 
spectroscopic information is reduced to photometry in several 
broad-band filters. Some ``color'' information is also available 
from large-scale surveys such as LAMOST \citep{2012RAA....12..723Z}. 
Quite a few studies have already been published that discuss the 
multi-wavelength results of stellar flares 
\citep[see a review by][]{2024LRSP...21....1K}. 
To summarize, chromospheric lines are indicators of solar-like 
activity, not just of flares. There is a correlation between these
activity indices and the occurrence of flares for active stars. The 
presence of a flare is indicated by the appearance of emission 
in these lines.

\section{Questions and a Path Forward}

Despite continuing efforts 
\citep[e.g.,][]{2025MNRAS.537..537O,2025arXiv251107036L}, many open 
questions remain about the spectrum of flare radiation, its origin, 
time evolution and frequency. Summarizing them:

{\bf (I)} Statistics of flares (power distribution, frequency) as a
function of spectral type and stellar age requires spectroscopic (or 
simultaneous multi-band UV-to-IR) monitoring of thousands of flare 
stars to characterize the SED of the flares; the total number and 
duration depends on the frequency of flares and the sensitivity of the 
future facility -- to ensure a statistically significant number of 
flares ($\sim$100s per T$_{eff}$/[Fe/H]/age category of stars to obtain 
reliable trends).

{\bf(II)} How dangerous are the flares to complex life and in 
particular -- to Earth's life? This study will benefit 
from a collaboration with biologists working on extremophiles; at this
time we cannot be more specific about the state of radiation biology 
and what specific questions will be relevant in the 2040s, but clearly 
this project will benefit from an interdisciplinary cooperation.

\section{Future Milestones, Synergies, Requirements, and Strategy}

The following considerations are based on the assumption that a 
facility similar to the WST
\citep[Wide Field Survey telescope;][]{2024arXiv240305398M} will 
become available at ESO in the 2040s. It is foreseen to have 30,000 
fibers and 9\,arcmin$^2$ integral field unit.

The ensemble studies of flares depend on assembling a large sample 
of (mostly late-type) stars, and on spectroscopic high-cadence 
monitoring. Stars with known habitable planets -- tens or hundreds -- 
are too few to provide ensemble information. They orbit bright stars
and can be followed-up with relatively small telescopes, on an 
individual basis. Instead, for the statistical study the large ensemble 
will have to include fainter stars with $V$$>$9\,mag, observe regions 
near the Milky Way plane or inner regions, and take advantage of a
high-multiplexity facility.
Sources for the sample, e.g. for M-stars, are: {\it Gaia} (DR4 expected 
in 2026), the ongoing surveys like SDSS (2000--) and planned like 4MOST 
(2026--). Multiband variability, including flare activity, will be 
assessed with ZTF (2018--) and the Rubin observatory (2025--). The 
{\it Euclid} mission (2023--) will provide precise space-based infrared 
photometry, that will allow to look into color excesses and redder 
companions.

This facility landscape of the 2040s implies a {\bf strategy} that combines 
two options: a continuous high-cadence monitoring of carefully selected 
late-type stars spanning the parameter space of interest and triggered 
follow-up observations for stars that exhibit a flare during 
the observations. The second option requires that the integration time 
is always split in at least two sub-integrations, and unlike the usual 
practice of template-completion-driven data reduction, in this case the 
data reduction is performed immediately upon completion and transfer of 
each sub-integration. Next, the reduction should not be longer than the 
time for the second sub-integration and its readout. If indications of 
a flare are found in the product spectrum, then the telescope should 
continue observing (if allowed by airmass and sky conditions) in the 
same configuration, until the flare is over. Therefore, the vast majority 
of the time WST will carry out its regular survey where different projects
share fibers in each configuration, and only in those rare occasions 
when a flare is detected during the observations, the regular short-term 
scheduling will be affected, albeit at the price of creating a flexible, 
fast and robust data flow system.

\section{Summary}

The flare characterization places the following 
{\bf requirements} considering a WST-like project:

{\bf (I)} A telescope larger than 4-m class is needed to obtain comparable 
statistics for fainter late-type stars and to measure the narrow 
chromospheric lines with short -- up to a minutes -- integrations to 
avoid ``smearing'' the short-lived flares.

{\bf (II)} Extreme multiplexity of $\sim$10$^4$ or higher.

{\bf (III)} A wide field of view of $\sim$1--3\,deg is needed for 
efficient observations.

{\bf (IV)} For reliable line profile fitting a resolving power of 
$\sim$65,000 is needed \citep{2015ApJ...813..125K}.

{\bf (V)} Finally, it is crucial to adopt flexible observing and reduction 
tools that allow implementation of the described observing strategy.

{\small
\bibliographystyle{apalike}
\setlength{\bibsep}{-2.0pt}
\bibliography{biblio_01}
}

\end{document}